\documentclass[pdflatex,sn-basic]{sn-jnl}

\usepackage{graphicx,times}             
\usepackage{natbib}
\usepackage{amssymb,amsmath}

\bibpunct{(}{)}{;}{a}{}{,}
\usepackage[]{hyperref}

\begin{document}

\title{Revisiting the infrared/X-ray correlation of GX 339$-$4 based on a jet model}

\author[1]{Chang-Yin Huang}\email{hcy@yangtzeu.edu.cn}
\author[2]{Yi Xie}\email{xieyi@jmu.edu.cn}

\affil[1]{School of Physics and Optoelectronic Engineering, Yangtze University, Jingzhou 434023, Hubei Province, China}
\affil[2]{School of Science, Jimei University, Xiamen 361021, Fujian Province, China}

\abstract{The infrared (IR)/X-ray correlation of GX 339$-$4 is investigated based on a jet model with a modification by linking the magnetic field at the jet base to the accretion rate of the inner accretion flow though the the equilibrium between magnetic pressure at horizon and the ram pressure of the accretion flow. The IR flux is attributed to the synchrotron radiation of the jet, and the X-ray flux is attributed to the advective dominated accretion flow (ADAF), synchrotron radiation of the jet and synchrotron self-Compton scattering (SSC) of the jet, respectively. We find that the observed IR/X-ray correlation with a break is well reproduced with the variation of the accretion rate if the X-ray flux originates from SSC of the jet. Either a conical ballistic jet with the magnetic field parallel to the jet axis or a conical adiabatic jet with an isotropic field can account for the correlation. The power-law index of the energy distribution of electrons $p\sim3$, the minimum Lorentz factor of the electrons $\gamma_{\rm min}\sim60$, the magnetic field $B_0\sim10^5\ {\rm G}$ and the jet radius $R_0\sim10^{10}\ {\rm cm}$ at the jet base are required for both the ballistic jet and the adiabatic jet. This study helps us clarify the complex interaction between the accretion and jet in GX 339$-$4, as well as the properties and geometric structure of the jet, laying the groundwork for exploring similar astrophysical systems.}


\maketitle

\section{Introduction}\label{sec1}

A black hole binary (BHB), consisting of a stellar mass black hole (BH) and a nondegenerate secondary star, is usually observed in an X-ray outburst lasting for weeks or months \citep{RM06}. During an outburst, different X-ray states have been classified, namely, low-hard state, high-soft state and intermediate state \citep{bell05,RM06}. An outburst begins in the low-hard state during which the luminosity is low and the spectrum is ``hard'' which can be described by a non-thermal power-law flux with typical photon index $\sim$1.6-1.7 \citep{RM06,bell10}. As the luminosity increases and the spectrum softens, the source enters the intermediate state with strong thermal and non-thermal components. The total spectrum is steep with typical photon index $\sim$2.4-2.5 \citep{RM06,bell10}. The duration of the intermediate state is usually short, but the source can move back and forth several times and shows fast and complex variability \citep{bell10}. Then the source enters the high-soft state, the luminosity of which is high and the spectrum is dominated by a thermal disk component. In the decaying phase of an outburst, the source moves back to the low-hard state through the intermediate state. It is widely believed that the high-soft state is well interpreted by the standard thin accretion disk \citep{SS73}, while the low-hard state can be successfully explained by the advective dominated accretion flow \citep[ADAF; also called radiatively inefficient accretion flow, RIAF,][]{NY95} or the outflowing corona models \citep[e.g.,][]{belo99,you21,DB24}. However, the intermediate state has not been well understood yet. 

Another markable feature of the low-hard state is the presence of jets which are collimated relativistic outflows ejected from the vicinity of the central BH and usually appear in pairs located on both sides of the accretion disk perpendicular to the disk surface. Jets are continuously observed in the hard state and they are switched off once the source enters the soft state, and are turned on again once the source return back to the hard state \citep{FBG04,bell10}. How the jet is generated and collimated is still an unresolved problem. The promising mechanisms include the BZ \citep{BZ77} and BP \citep{BP82} models which invoke large-scale magnetic fields to extract the energy of the rotational BH and disk, respectively. Jets are mainly detected in the radio to infrared band usually with flat or slightly inverted spectra (the observed flux density $S_{\nu}\propto\nu^{\alpha}$ with $\alpha\gtrsim0$), which can be explained by the superposition of synchrotron emission from different segments of the jet \citep{BK79}. In the model of  \cite{BK79}, the jet is assumed to have a conical geometry, a constant bulk velocity along the jet axis, and magnetic fields perpendicular to the jet axis. And the jet is assumed to be in a steady state without considering the adiabatic cooling and the radiative cooling of electrons. Several authors have developed the \cite{BK79} model by including the energy losses of the electrons \citep[e.g.,][]{mars80,koni81,reyn82,HJ88,SK94,GM98,kais06,PC09}. \cite{kais06} constructed a model for the synchrotron emission of partially self-absorbed jets including the effects of energy losses of the relativistic electrons due to the synchrotron process itself and the adiabatic expansion of the jet flow. Both the ballistic and adiabatic jet models were considered with the decay of the magnetic field and the flow material along the jet. Different magnetic field configurations are considered, i.e., perpendicular to the jet axis, parallel to the jet axis and isotropic fields. The ballistic jets, initially highly over pressured with respect to their environments, expand unimpeded with the jet material travelling along straight trajectories and produce flat radio spectra naturally with perpendicular magnetic fields. The adiabatic jets, confined by the surrounding material, dissipate energy to the external gas during the adiabatic expansion and can also produce flat radio spectra with a specific jet geometry. The results of the \cite{kais06} model are in consistent with the observational data of the BHB Cygnus X-1.

Correlations between multi-band radiations have been discovered for BHBs, e.g., the radio/X-ray correlation with the X-ray luminosity $L_{\rm X}$ being proportional to the radio luminosity $L_{\rm R}$ by the form $L_{\rm X}\propto L_{\rm R}^{0.6-0.7}$\citep{corb03,corb13,GFP03} and the infrared (IR)/X-ray correlation \citep{homa05,russ06,russ07,C09}. \cite{russ06} analyzed quasi-simultaneous optical/infrared (OIR)-X-ray observations of 33 X-ray binaries and found a global correlation between OIR and X-ray luminosity for BHBs in the hard state, of the form $L_{\rm OIR}\propto L_{\rm X}^{0.6}$. The authors explained the OIR/X-ray correlation based on the empirical relationships of the accretion-jet model, and found that the jet contributes about 90\% of the near-infrared radiation in the bright hard state. It is generally believed that the infrared luminosity mainly comes from synchrotron radiation of the jet, while X-ray luminosity may originate from the Comptonizing corona/ADAF, synchrotron radiation of the jet, or synchrotron self-Compton scattering (SSC)  \citep{mark05}. \cite{C09} (hereafter C09) analyzed the broad-band observations of four outbursts of the BHB GX 339-4 and found a tight correlation between the OIR luminosity and X-ray luminosity with the presence of a break in the IR/X-ray correlation in the hard state. The IR H-band flux density $F_{\rm IR}$ is proportional to 3-9 keV X-ray flux $F_{\rm X}$ with the form $F_{\rm IR}\propto F_{\rm X}^b$, where $b\sim0.68$ at low luminosity and $b\sim0.48$ at high luminosity. C09 explained the IR/X-ray correlation qualitatively based on the empirical relationship between SSC luminosity $L_{\rm X}$ and the jet power $Q_{\rm jet}$, i.e., $L_{\rm X}\propto Q_{\rm jet}^{11/4}$, as well as the relation between the monochromatic luminosity $L_{\nu}$ and the jet power, i.e., $L_{\nu}\propto Q_{\rm jet}^{17/12-2\alpha/3}$, from the standard conical jet model, where $\alpha$ is the jet spectral index. The break in the IR/X-ray correlation can be interpreted by the transition from the optically thick synchrotron radiation at H band to the optically thin one as X-ray luminosity decreases.

Although the IR/X-ray correlation of GX 339$-$4 was well interpreted by the qualitative analysis in C09, however, there is a lack of quantitative calculations and simulations on the IR/X-ray correlation in the literature. In this paper, we try to calculate quantitatively the IR/X-ray correlation based on the \cite{kais06} jet model considering different X-ray origins, i.e., synchrotron radiation from the jet, SSC of the jet and the Comptonizing corona. We compare our modeled results with the observed correlation given in C09 and discuss the jet parameters of GX339$-$4.

\section{The model} \label{sec2}

\subsection{Synchrotron radiation from jets}

It is usually assumed that electrons in a jet have a power-law energy distribution of the form
\begin{equation}\label{eq1}
N(\gamma){\rm d}\gamma=\kappa\gamma^{-p}{\rm d}\gamma,
\end{equation}
where $\gamma$ is the Lorentz factor of the electron, $N$ is the number density of electrons with Lorentz factors between $\gamma$ and $\gamma+{\rm d}\gamma$, and $\kappa$ is a scaling independent of $\gamma$. The emission coefficient $j_\nu$ and the absorption coefficient $\alpha_\nu$ of synchrotron radiation from the power-law electron distribution can be expressed as functions of the magnetic field $B$ and frequency $\nu$ respectively as \citep{RL79}
\begin{equation}\label{eq2}
j_\nu=A_p\kappa B^\frac{p+1}{2}\nu^{-\frac{p-1}{2}} \ {\rm erg\ cm^{-3}\ s^{-1}\ Hz^{-1}\ ster^{-1}}
\end{equation}
and
\begin{equation}\label{eq3}
\alpha_\nu=C_p\kappa B^\frac{p+2}{2}\nu^{-\frac{p+4}{2}} \ {\rm cm^{-1}},
\end{equation}
where
\begin{equation*}
A_p=\frac{\sqrt{3}e^3}{4\pi m_{\rm e}c^2(p+1)}\left(\frac{2\pi m_{\rm e}c}{3e}\right)^{-\frac{p-1}{2}}\Gamma\left(\frac{p}{4}+\frac{19}{12}\right)\Gamma\left(\frac{p}{4}-\frac{1}{12}\right),
\end{equation*}
\begin{equation*}
C_p=\frac{\sqrt{3}e^3}{8\pi m_{\rm e}^2c^2}\left(\frac{2\pi m_{\rm e}c}{3e}\right)^{-\frac{p}{2}}\Gamma\left(\frac{p}{4}+\frac{1}{6}\right)\Gamma\left(\frac{p}{4}+\frac{11}{12}\right),
\end{equation*}
where $e$ is the charge of the electron, $m_{\rm e}$ is the mass of the electron, $c$ is the speed of light, and $\Gamma(x)$ is the gamma function of argument $x$.

Following \cite{kais06}, we assume that the jet radius $R$, the magnetic field $B$ and $\kappa$ are all power-law functions of the jet hight $z$, i.e.,
\begin{equation}\label{eq4}
R=R_0\left(\frac{z}{z_0}\right)^{a_1}, \ B=B(z_0)\left(\frac{z}{z_0}\right)^{-a_2}, \ \kappa=\kappa_0\left(\frac{z}{z_0}\right)^{-a_3},
\end{equation}
where $0\leqslant a_1\leqslant1$ describing the shape of the jet and $a_1\leqslant a_2\leqslant2a_1$ describing the geometry of the magnetic field. We have $a_2=a_1$ for a perpendicular field as to the jet axis, $a_2=2a_1$ for a parallel field and $a_2=4a_1/3$ for an isotropic field. $z_0$ is an arbitrary position along the jet axis and we choose it as the height where the synchrotron radiation starts. For ballistic jets we have $a_3=2a_1$ based on the particle conservation,  and for adiabatic jets we have $a_3=(4+2p) a_1/3$ by including the energy losses due to the jet expansion \citep{kais06}.  $\kappa_0$ in equation (4) can be estimated by assuming that the energy densities of the magnetic field and of the relativistic electrons are in equipartition at $z_0$ \citep{kais06}. Therefore we have
\begin{equation*}
\frac{B(z_0)^2}{8\pi}=\int^{\gamma_{\rm max}}_{\gamma_{\rm min}}(\gamma m_{\rm e}c^2)\kappa_0\gamma^{-p}{\rm d}\gamma=\left\{ \begin{array}
{r@{\quad \quad}l}
\frac{\kappa_0 m_{\rm e}c^2}{p-2}(\gamma_{\rm min}^{2-p}-\gamma_{\rm max}^{2-p}),  &  p\neq 2 \\ \kappa_0 m_{\rm e}c^2\ln{\frac{\gamma_{\rm max}}{\gamma_{\rm min}}}, &  p=2
\end{array} \right.
\end{equation*}
and
\begin{equation}\label{eq5}
\kappa_0=\frac{f_p B_0^2\dot{m}}{8\pi m_{\rm e}c^2},
\end{equation}
where
\begin{equation*}
f_p=\left\{ \begin{array}
{r@{\quad \quad}l}
\frac{p-2}{\gamma_{\rm min}^{2-p}-\gamma_{\rm max}^{2-p}},  &  p\neq 2 \\ \frac{1}{\ln{\frac{\gamma_{\rm max}}{\gamma_{\rm min}}}}. &  p=2
\end{array} \right.
\end{equation*}
In equation \eqref{eq5}, we have assumed $B(z_0 )=B_0 \dot{m}^{1/2}$ based on the assumption that the magnetic field at $z_0$ is proportional to that at the BH horizon, which is proportional to $\dot{m}^{1/2}$ based on the equilibrium between magnetic pressure at horizon and the ram pressure of the innermost parts of the accretion flow \citep{MSL97}, where $\dot{m}$ is the dimensionless accretion rate in Eddington ratio and $B_0$ is the scale factor for the magnetic field at $z_0$. 

The monochromatic intensity from a segment of the jet is
\begin{equation*}
I_\nu=\int^R_0 j_\nu {\rm e}^{-\alpha_\nu(R-s)}{\rm d}s=\frac{j_\nu}{\alpha_\nu}(1-{\rm e}^{-\alpha_\nu R}).
\end{equation*}
Adding up the intensities from all the segments of the radiation zone, we obtain the observed flux density (the Doppler factor is ignored for simplicity and the jet axis is assumed perpendicular to the line of sight)
\begin{equation}\label{eq6}
F_\nu=\int^{z_{\rm m}}_{z_0}\frac{2R I_\nu{\rm d}z}{d^2}=\frac{2}{d^2}\int^{z_{\rm m}}_{z_0}\frac{j_\nu R}{\alpha_\nu}(1-{\rm e}^{-\alpha_\nu R}){\rm d}z,
\end{equation}
where $z_{\rm m}$ is the maximum height of the radiation zone, and $d$ is the distance from the source to the earth. 

\subsection{SSC luminosity in jets}

For a power-law distribution of electrons, the spectrum of the inverse-Compton  scattering is given as \citep{jone68,BG85}
\begin{equation}\label{eq7}
L_\nu=h\nu\int{\rm d}V\int{\rm d}\nu' n(\nu')\frac{3\sigma_{\rm T}\kappa c}{4\nu}2^p\left(\frac{\nu'}{\nu}\right)^{(p-1)/2}F(\nu,\nu'),
\end{equation}
where
\begin{equation*}
F(\nu,\nu')=\int^{q_{\rm max}}_{q_{\rm min}}\frac{q^{(p-1)/2}[2q\ln{q}+(1+2q)(1-q)+2(1-q)fq]}{[1+\sqrt{f q/(1+f q)}]^{p+2}(1+f q)^{(p+3)/2}}{\rm d}q,
\end{equation*}
\begin{equation*}
f=\left(\frac{h}{m_{\rm e} c^2}\right)^2\nu\nu',
\end{equation*}
\begin{equation*}
q_{\rm min}=\frac{\nu}{\nu'4\gamma^2_{\rm max}[1-h\nu/(\gamma_{\rm max}m_{\rm e}c^2)]}, \ \ \ \ \ \ (0\lt q_{\rm min}\le 1)
\end{equation*}
\begin{equation*}
q_{\rm max}=\frac{\nu}{\nu'4\gamma^2_{\rm min}[1-h\nu/(\gamma_{\rm min}m_{\rm e}c^2)]}, \ \ \ \ \ \ (0\lt q_{\rm max}\le 1)
\end{equation*}
where $\nu'$ is the frequency of the incident photons, $\nu$ is the frequency of the scattered photons, ${\rm d}V=\pi R^2 {\rm d}z$ is the infinitesimal volume of the scattering region, $\sigma_T$ is the Thomson cross section and $h$ is the Planck's constant. $n(\nu')$ in equation \eqref{eq7} is the number density of the incident photons which are provided by synchrotron radiation in the jet. For simplicity, we estimated as
\begin{equation}\label{eq8}
n(\nu')=\frac{4\pi I_{\nu'}}{h\nu' c}=\frac{4\pi}{h c}\frac{A_p}{C_p}B^{-1/2}{\nu'}^{3/2}(1-{\rm e}^{-\alpha_{\nu' R}}).
\end{equation}
Substituting equations \eqref{eq8} and  \eqref{eq4} into equation \eqref{eq7} , integrating from 3 to 9 keV and dividing by $4\pi d^2$, we obtain the 3-9 keV SSC flux
\begin{equation}\label{eq9}
\begin{split}
F_{\rm SSC,3-9\ keV} &= \frac{3 \sigma_{\rm T}2^p f_p A_p}{32 m_{\rm e}c^2d^2C_p} z_0^{-a_9}R_0^2 B_0^{3/2}\dot{m}^{3/4}\int^{\rm 9\ keV/h}_{\rm 3\ keV/h}\nu^{-\frac{p-1}{2}} \\
& \int^{z_{\rm m}}_{z_{\rm 0}}z^{a_9}\int^\nu_{\nu'_{\rm min}}{\nu'}^\frac{p+2}{2}(1-{\rm e}^{-\alpha_{\nu'}R})F(\nu,\nu'){\rm d}\nu'{\rm d}z{\rm d}\nu,
\end{split}
\end{equation}
where
\begin{equation*}
\nu'_{\rm min}=\frac{\nu}{4\gamma^2_{\rm min}[1-h\nu/(\gamma_{\rm min}m_{\rm e}c^2)]}. 
\end{equation*}

\subsection{X-ray flux from the Comptonizing corona}

In the low-hard state of BHBs, the radiation efficiency is usually thought to be low and the spectra are successfully explained by the corona/ADAF model, in which the X-ray luminosity is proportional to the square of the accretion rate \citep[$L_{\rm X}\propto\dot{m}^2$, e.g.,][] {maha97}. \cite{kord06} measured the accretion rate of the hard state BHBs by the radio emission of the jet and obtained a relationship between the accretion rate and 2-10 keV X-ray luminosity using the radio/X-ray correlation. We use this relation to estimate the 2-10 keV X-ray luminosity:
\begin{equation}\label{eq10}
L_{\rm 2-10\ keV}=4.3\times10^{37}m^{1.14}\dot{m}^2\ {\rm erg\ {\rm s}^{-1}}, 
\end{equation}
where $m\equiv M/M_\odot$, $M$ is the BH mass and $M_\odot$ is the solar mass. To compare with the observed IR/X-ray correlation given in C09, we calculated the 3-9 keV luminosity based on the 2-10 keV luminosity assuming a power-law spectrum ($L_E\propto E^{-\Gamma}$) with typical spectral index $\Gamma\sim0.6$ in the hard state. The 3-9 keV X-ray flux
\begin{equation}\label{eq11}
F_{\rm 3-9\ keV}=2.5\times10^{36}d^{-2}m^{1.14}\dot{m}^2\ {\rm erg\ cm^{-2}\ s^{-1}}. 
\end{equation}

\section{A qualitative analysis of the IR/X-ray correlation}

To calculate the synchrotron flux density, it is convenient to rewrite \eqref{eq6}  into the form as an integration of the optical depth $\tau$:
\begin{equation}\label{eq12}
F_\nu=\frac{A_p C_p^{a_6}f_p^{a_6+1}z_0 R_0^{a_6+2}B_0^{2a_7}\dot{m}^{a_7}\nu^{a_8}}{4\pi d^2 (m_{\rm e} c^2)^{a_6+1} a_5}\int^{\tau_{\rm m}}_{\tau_0}\tau^{-a_6-2}(1-{\rm e}^{-\tau}){\rm d}\tau,
\end{equation}
by using 
\begin{equation}\label{eq13}
\tau\equiv\alpha_\nu R=\frac{f_p C_p R_0}{8\pi m_{\rm e}c^2}B_0^{\frac{p+6}{2}}z_0^{-a_5}\dot{m}^{\frac{p+6}{4}}\nu^{-\frac{p+4}{2}}z^{a_5},
\end{equation}
where $a_5=-(p+2)a_2/2+a_1-a_3$, $a_6=[(p+1)a_2/2+a_3-2a_1-1]/a_5$, $a_7=(p+6)a_6/4+(p+5)/4$, $a_8=-(p+4)a_6/2-(p-1)/2$. $\tau_0$ and $\tau_{\rm m}$ are the optical depths at $z_0$ and $z_{\rm m}$ respectively.

When the accretion rate $\dot{m}$ is low, i.e., if $\tau_0\lt 1$, the integrand in equation \eqref{eq12} can be simplified as $\tau^{-a_6-1}$, therefore
\begin{equation}\label{eq14}
F_\nu\propto\dot{m}^{a_7-\frac{(p+6)a_6}{4}}=\dot{m}^{\frac{p+5}{4}}
\end{equation}
based on equations \eqref{eq12} and \eqref{eq13}. When the accretion rate $\dot{m}$ is high, part of the jet becomes optically thick, i.e., $\tau_0\gg1$ and $\tau_{\rm m}\sim0$, then the integration in equation \eqref{eq12} can be approximately as
\begin{equation*}
\int^{\tau_{\rm m}}_{\tau_0}\tau^{-a_6-2}(1-{\rm e}^{-\tau}){\rm d}\tau\approx \int^1_\infty\tau^{-a_6-2}{\rm d}\tau+\int^0_1\tau^{-a_6-1}{\rm d}\tau=\frac{1}{a_6(a_6+1)},
\end{equation*}
and we have
\begin{equation}\label{eq15}
F_\nu\propto\dot{m}^{a_7}.
\end{equation}

We can obtain the IR/X-ray correlation with the variation of $\dot{m}$. If the IR flux originates from the synchrotron radiation of the jet and X-ray flux originates from the Comptonizing corona, the IR/X-ray correlation $F_{\rm IR}\propto F_{\rm X}^b$ predicted by the model with the index $b$ changing from $(p+5)/8$ to $a_7/2$ with the increasing $\dot{m}$ based on equations \eqref{eq14}, \eqref{eq15}  and \eqref{eq11}. For $2\leqslant p\leqslant3$, we have $0.875\leqslant(p+5)/8\leqslant1$ which is larger than the index $\sim0.68$ of the steeper branch of GX339$-$4 given in C09. For a conical adiabatic jet, we have $0.46\leqslant a_7/2\leqslant0.56 $ with magnetic fields perpendicular to the jet axis ($a_2=1$) and $0.40\leqslant a_7/2\leqslant0.49$ with the isotropic magnetic fields, which are both in accordance with the index $\sim0.48$ of the observed gentler branch.  For a conical adiabatic jet with parallel magnetic fields, a conical ballistic jet and a parabolic jet, the modeled correlation is not consistent with the observed gentler branch.

If both the IR flux and X-ray flux originate from the synchrotron radiation of the jet, we have $F_{\rm IR}\propto F_{\rm X}$ at low accretion rate since the jet is optically thin for both IR and X-ray bands. This is too steep for the steeper branch. As $\dot{m}$ increases, if the IR band (especially the H-band) becomes optically thick and the X-ray band is still optically thin, then we have $F_{\rm IR}\propto F_{\rm X}^b$ with $b=4a_7/(p+5)$ based on equations \eqref{eq14} and \eqref{eq15}. For a conical ballistic jet, we have $0.44\leqslant b\leqslant0.54$ with $ a_2=2$, and for a conical adiabatic jet, we have $0.46\leqslant b\leqslant0.64$ with $a_2=1$ and $0.40\leqslant b\leqslant0.56$ with $a_2=4/3$, which are all in accordance with the index $\sim0.48$ of the observed gentler branch. The correlation of the parabolic jet is not consistent with the observed gentler branch.

If the synchrotron radiation and SSC of the jet are responsible for the IR and X-ray fluxes respectively, we need to analyze the correlation between SSC flux and accretion rate $\dot{m}$ based on equation \eqref{eq9}. If all the synchrotron radiation photons scattered into the energy range 3-9 keV are optically thin ($\alpha_{\nu'}R\ll 1$), the factor $1-{\rm e}^{\alpha_{\nu'}R}$ in equation \eqref{eq9} is approximately $\alpha_{\nu'}R$, and we have
\begin{equation}\label{eq16}
F_{\rm SSC,3-9\ keV} \propto\dot{m}^{\frac{3}{4}+\frac{p+6}{4}}=\dot{m}^{\frac{p+9}{4}}.
\end{equation}
 If part of the synchrotron radiation photons becomes optically thick, it is convenient to rewrite equation \eqref{eq9} into the form as an integration of the optical depth  $\tau'\equiv\alpha_{\nu'}R$. 
We find that $F(\nu,\nu')$ in equation \eqref{eq9} can be approximated by a fitting function $F(\nu,\nu')\propto(\nu'/\nu)^{-1.7}$.  Substituting it into equation \eqref{eq9} we have
\begin{equation}\label{eq17}
F_{\rm SSC,3-9\ keV} \propto\dot{m}^{\frac{3}{4}+\frac{(p+6)(5p+3)}{20(p+4)}}\int^{\tau(\nu)}_{\tau'(\nu'_{\rm min})}\tau'^{-\frac{10p+23}{5p+20}}(1-\rm{e}^{-\tau'}){\rm d}\tau'.
\end{equation}
The integration in equation \eqref{eq17} is independent of $\dot{m}$ since the upper bound $\tau(\nu)\sim0$ and the lower bound $\tau'(\nu'_{\rm min})\gg 1$ which can be replaced by $\infty$.
Therefore, we have
\begin{equation}\label{eq18}
F_{\rm SSC,3-9\ keV} \propto\dot{m}^{\frac{3}{4}+\frac{(p+6)(5p+3)}{20(p+4)}}.
\end{equation}
Then we have $F_{\rm IR}\propto F_{\rm SSC,3-9\ keV}^b$ with $b=(p+5)/(p+9)$ at low accretion rates from equations \eqref{eq14} and \eqref{eq16}, and with $b=20(p+4)a_7/(5p^2+48p+78)$ at high accretion rates from equations \eqref{eq15} and \eqref{eq18}. We have $0.64\leqslant b\leqslant0.67$ with $2\leqslant p\leqslant3$ at low accretion rates. At high accretion rates, for the conical ballistic jet, we have $0.46\leqslant b\leqslant0.59$ with $a_2=2$. For the conical adiabatic jet, we have $0.48\leqslant b\leqslant0.69$ and $0.35\leqslant b\leqslant0.50$ with $a_2=1$ and $a_2=2$, respectively. The correlation of the parabolic jet is not consistent with the gentler branch of the observed correlation. Either a conical ballistic jet with parallel magnetic fields or a conical adiabatic jet can account for the break of the IR/X-ray correlation, through the variation of the accretion rate $\dot{m} $ only.

In summary, If the X-ray flux originates from the Comptonizing corona or the synchrotron radiation of the jet, the predicted IR/X-ray correlation is consistent with observations at high luminosity. However, at low luminosity, the predicted correlation is too steep for the steeper branch of the observed correlation. If the IR and X-ray fluxes originate from the synchrotron radiation and SSC of the jet respectively, then the predicted correlation is consistent with observations at both high and low luminosities.

\section{Numerical results} \label{sec4}

We calculate numerically the $F_{\nu_{\rm H}}$-$F_{\rm 3-9 keV}$ correlation based on equations \eqref{eq6}, \eqref{eq9} and \eqref{eq11} with the variation of the accretion rate $\dot{m}$, where we substitutie $\nu=\nu_{\rm H}=1.8\times10^{14}\ {\rm Hz}$ into equation \eqref{eq6}. We consider different X-ray origins, i.e., the synchrotron radiation of the jet, SSC of the jet and the Comptonizing corona. Since the values of $z_{\rm m}$ and $\gamma_{\rm max}$ are not sensitive to the correlation for all these X-ray origins, we take $z_{\rm m}=10^5 z_0$ and $\gamma_{\rm max}=10^5$ in our calculations. The black hole mass $m=5.8$ \citep{hyne03} and the distance $d=8\ {\rm kpc}$ \citep{zdzi04} are taken, respectively. The free parameters include $p$, $a_1$, $a_2$, $a_3$, $z_0$, $B_0$, $R_0$ and $\gamma_{\rm min}$.

\begin{figure}
\centering
\includegraphics[width=0.7\columnwidth]{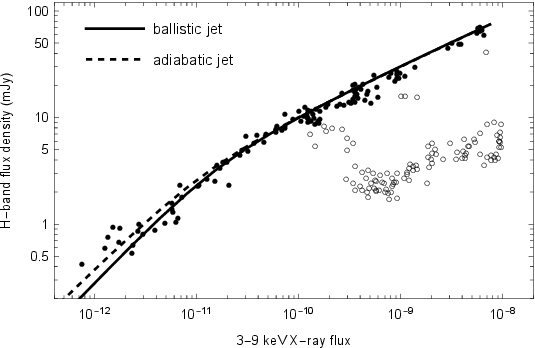}
\caption{The comparison of the modeled IR/X-ray correlation with the corona X-ray origin to the observed one of GX339$-$4. The observational data (filled circles denote the hard state, and empty circles denote the soft and intermediate states) are taken from C09 without error bars. The solid line is the modeled correlation for the ballistic jet with $p=3$, $a_1=1$, $a_2=a_3=2a_1$, $z_0=5.14\times10^6\ {\rm cm}$, $B_0=2\times10^5 \ {\rm G}$, $R_0=7.5\times10^9 \ {\rm cm}$ and $\gamma_{\rm min}=10$. The dashed line is the modeled correlation for the adiabatic jet with $p=2.5$, $a_1=1$, $a_2=4a_1/3$, $a_3=(4+2p)a_1/3$, $z_0=5.14\times10^6\ {\rm cm}$, $B_0=2.5\times10^5 \ {\rm G}$, $R_0=4.5\times10^9 \ {\rm cm}$ and $\gamma_{\rm min}=10$.}
\label{fig:1}
\end{figure}

In Figure \ref{fig:1}, we compare the calculated correlation with the corona X-ray origin to the observed one of GX 339$-$4 given in C09. We find that the smaller the $z_0$ value, the better the theoretical curve matches the observed one. So in our calculations, we take the value of $z_0$ as the theoretical limit at which the jet extends closest to the black hole, i.e., about $6GM/c^2$ as suggested by \cite{kais06}. For GX 339$-$4, the lower limit of $z_0$ is $5.14\times10^6  {\rm cm}$. We also find that the modeled correlation is not very sensitive to the value of $\gamma_{\rm min}$, and the effect of $\gamma_{\rm min}$ on the correlation can be achieved by adjusting the values of $B_0$ and $R_0$ to obtain the same results. We take $\gamma_{\rm min}=10$ following \cite{BK79}. As shown in Figure 1, the correlations obtained from the ballistic jet and the adiabatic jet are both in good agreement with the observed gentler branch at high luminosity. $a_1=1$ and $a_2=2a_1$ are required for the ballistic jet suggesting a conical jet with the magnetic field parallel to the jet axis.  $a_1=1$ and $a_2=4a_1/3$ are required for the adiabatic jet suggesting a conical jet with an isotropic magnetic field. Both models suggest the magnetic field $B_0\sim10^5 \ {\rm G}$ and the jet radius $R_0\sim10^9 \ {\rm cm}$ at the jet base. The ballistic jet prefers a larger $p$ value ($p=3$),  while the adiabatic jet prefers a smaller $p$ value ($p=2.5$), which may be due to the energy loss during the adiabatic expansion.

\begin{figure}
\centering
\includegraphics[width=0.7\columnwidth]{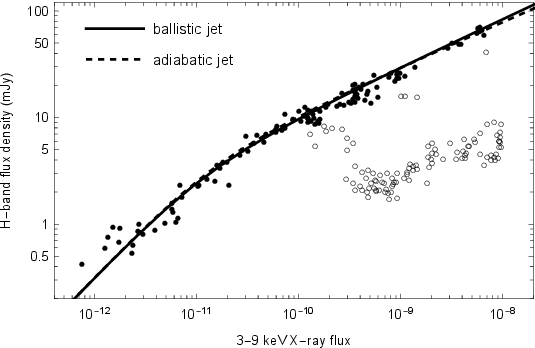}
\caption{The comparison of the modeled IR/X-ray correlation with the origin of synchrotron radiation in X-rays to the observed one of GX339$-$4. The observational data points are the same as Fig. 1. The solid line is the modeled correlation for the ballistic jet with $p=2.9$, $a_1=1$, $a_2=a_3=2a_1$, $z_0=2\times10^7\ {\rm cm}$, $B_0=5\times10^5 \ {\rm G}$, $R_0=1.7\times10^9 \ {\rm cm}$ and $\gamma_{\rm min}=10$. The dashed line is the modeled correlation for the adiabatic jet with $p=2.9$, $a_1=1$, $a_2=4a_1/3$, $a_3=(4+2p)a_1/3$, $z_0=2\times10^7\ {\rm cm}$, $B_0=5\times10^5 \ {\rm G}$, $R_0=2\times10^9 \ {\rm cm}$ and $\gamma_{\rm min}=10$.}
\label{fig:2}
\end{figure}

\begin{figure}
\centering
\includegraphics[width=0.7\columnwidth]{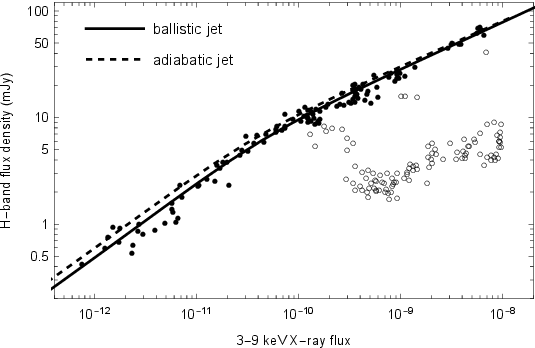}
\caption{The comparison of the modeled IR/X-ray correlation with the SSC X-ray origin to the observed one of GX339$-$4. The observational data points are the same as Fig. 1.  The solid line is the modeled correlation for the ballistic jet with $p=3$, $a_1=1$, $a_2=a_3=2a_1$, $z_0=5.14\times10^6\ {\rm cm}$, $B_0=1.5\times10^5 \ {\rm G}$, $R_0=1.0\times10^{10} \ {\rm cm}$ and $\gamma_{\rm min}=60$. The dashed line is the modeled correlation for the adiabatic jet with $p=3$, $a_1=1$, $a_2=4a_1/3$, $a_3=(4+2p)a_1/3$, $z_0=5.14\times10^6\ {\rm cm}$, $B_0=2\times10^5 \ {\rm G}$, $R_0=1.0\times10^{10} \ {\rm cm}$ and $\gamma_{\rm min}=55$.}
\label{fig:3}
\end{figure}

In Figure \ref{fig:2}, we compare the calculated correlation with X-rays coming from the synchrotron radiation of the jet to the observed one of GX 339$-$4. The calculated correlations are similar with those shown in Figure \ref{fig:1}, which are both in good agreement with the observed gentler branch at high luminosity and are too steep for the steeper branch at low luminosity. A conical ballistic jet with a parallel magnetic field or a conical adiabatic jet with an isotropic magnetic field are also suggested. $B_0\sim10^5 \ {\rm G}$ and $R_0\sim10^9 \ {\rm cm}$ are also required. $p=2.9$ and $z_0=2\times10^7\ {\rm cm}$ are both preferred for the ballistic jet and the adiabatic jet.

In Figure \ref{fig:3}, we compare the calculated correlation with the SSC X-ray origin to the observed one of GX 339$-$4. We take the lower limit value of $z_0$, $5.14\times10^6\ {\rm cm}$, since the theoretical curve matches the observed one better for a smaller $z_0$ value. We find that a larger value of $p$ makes the steeper branch more steep and the gentler branch more gentle. To fit the steeper branch, a larger $p$ value (we take $p=3$) is needed. The value of $B_0$ does not affect the slope and height of the curve, but it can affect the length of the curve. To cover all data points, $B_0\sim10^5 \ {\rm G}$ is needed for $0.001\lesssim\dot{m}\lesssim0.3$. The values of $R_0$ and $\gamma_{\rm min}$ do not affect the steeper branch. A larger $R_0$ value makes the gentler branch more gentle, while a larger $\gamma_{\rm min}$ value makes the gentler branch less gentle. For the above given values of $p$ and $B_0$, we find that $R_0\sim10^{10} \ {\rm cm}$ and $\gamma_{\rm min}\sim60$ are required for both the ballistic jet and the adiabatic jet. The calculated correlations for the two types of jets are both in good agreement with the observed one. The ballistic jet prefers a parallel magnetic field, while the adiabatic jet prefers an isotropic field. The conical shape is needed for both.

\section{Discussions and conclusions} \label{sec:conclusions}

We calculated the IR/X-ray correlation based on the \cite{kais06}  jet model with a modification by 
linking the magnetic field at the jet base to the accretion rate of the inner accretion flow though the the equilibrium between magnetic pressure at horizon and the ram pressure of the innermost parts of the accretion flow. With the variation of the accretion rate $\dot{m}$, we obtain the $F_{\nu_{\rm H}}$-$F_{\rm 3-9 keV}$ correlation with the H-band flux coming from the synchrotron radiation of the jet and 3-9 keV X-ray flux coming from the Comptonizing corona, synchrotron radiation of the jet and SSC of the jet, respectively. We compared the modeled IR/X-ray correlation with the observed one of GX 339$-$4 given in C09. The correlation with a corona or synchrotron origin of X-rays is consistent with the observed gentle branch at high luminosity, but is too steep for the steeper branch at low luminosity. The correlation with SSC origin of X-rays matches the observed one well both at high and low luminosities if the jet is conical, and the magnetic field is parallel to the jet axis for the ballistic jet and is isotropic for the adiabatic jet, respectively. $p=3$, $B_0\sim10^5\ {\rm G}$, $R_0\sim10^{10}\ {\rm cm}$ and $\gamma_{\rm min}\sim60$ are required for both the ballistic jet and the adiabatic jet.

Our results confirm the analysis of C09 that the IR/X-ray correlation can be interpreted by the SSC origin of X-rays and the break is caused by the transition from the optically thick synchrotron radiation at H band to the optically thin one as the X-ray luminosity decreases. When the X-ray luminosity is high, and therefore the accretion rate $\dot{m}$ is large, the jet is optically thick according to equation \eqref{eq13} since the optical depth $\tau$ is positively correlated to $\dot{m}$. When $\dot{m}$ is low, then $\tau$ is small, corresponding to optically thin jets.  In fact, the integrand in equation (12) is a broken power-law function of $\tau$ with the break at $\tau=1$ and therefore the synchrotron flux density is also a broken power-law function of $\dot{m}$ with the break at $\tau_0\sim1$. 
Substituting $\tau_0=1$ into equation (13), we obtain the accretion rate at the break
\begin{equation}\label{eq19}
\begin{split}
\dot{m}_{\rm b} &= \left(\frac{f_p C_p R_0}{8\pi m_{\rm e}c^2}\right)^{-\frac{4}{p+6}}\nu_{\rm H}^\frac{2(p+4)}{p+6}B_0^{-2}  \\
&\approx0.05\left(\frac{\gamma_{\rm min}}{10}\right)^{-\frac{4}{9}}\left(\frac{R_0}{10^{10}{\rm\ cm}}\right)^{-\frac{4}{9}}\left(\frac{B_0}{10^5{\rm\ G}}\right)^{-2}
\end{split}
\end{equation}
for $p=3$. When $\dot{m}\lesssim\dot{m}_{\rm b}$, the whole jet in H-band is optically thin, and the steeper branch with index $\sim0.68$ is produced. When $\dot{m}\gtrsim\dot{m}_{\rm b}$ leading to $\tau_{\nu_{\rm H}}(z_0)>1$ and $\tau_{\nu_{\rm H}}(z_{\rm m})<1$, i.e., part of the jet becomes optically thick and the downstream is still optically thin, then the gentler branch with index  $\sim0.48$ is produced. 
The corresponding 3-9 keV X-ray luminosities at the break for the corona, synchrotron and SSC origins are respectively (based on equations \eqref{eq11}, \eqref{eq9}, \eqref{eq12}):
\begin{equation}\label{eq20}
L_{\rm cor,b}\approx9.1\times10^{34}m^{1.14}\left(\frac{\gamma_{\rm min}}{10}\right)^{-\frac{8}{9}}\left(\frac{R_0}{10^{10}{\rm\ cm}}\right)^{-\frac{8}{9}}\left(\frac{B_0}{10^5{\rm\ G}}\right)^{-4} {\rm erg \ s^{-1}},
\end{equation}
\begin{equation}\label{eq21}
L_{\rm syn,b}\approx1.65\times10^{34}\left(\frac{\gamma_{\rm min}}{10}\right)^{\frac{1}{9}}\left(\frac{R_0}{10^{10}{\rm\ cm}}\right)^{\frac{10}{9}}\frac{z_0}{10^7{\rm\ cm}}\ {\rm erg \ s^{-1}},
\end{equation}
\begin{equation}\label{eq22}
L_{\rm SSC,b}\approx1.62\times10^{34}\left(\frac{\gamma_{\rm min}}{10}\right)^{\frac{61}{15}}\left(\frac{R_0}{10^{10}{\rm\ cm}}\right)^{\frac{5}{3}}\frac{z_0}{10^7{\rm\ cm}}\ {\rm erg \ s^{-1}},
\end{equation}
for $p=3$, $a_1=1$, $a_2=a_3=2$. Therefore, the model roughly predicts a break luminosity of $\sim10^{34-35} \ {\rm erg\ s^{-1}}$ with typical  jet parameters, corresponding to a flux of $\sim10^{-12}-10^{-11} \  {\rm erg\ cm^{-2}\ s^{-1}}$ for GX 339$-$4, which is consistent with the observed break flux in C09.

In this paper, we provide a possible explanation for the break of the IR/X correlation in GX 339$-$4. The observed correlation can be reproduced with the variation of the accretion rate only. However, \cite{russ13} investigated the jet spectral breaks of several BHBs and found that there is no clear empirical trend between jet break frequency and luminosity within the hard state. This may indicate that the accretion rate alone is not enough to describe the break of the IR/X correlation. Actually, other jet parameters, e.g., the magnetic field $B_0$, jet radius $R_0$ and the energy distribution index $p$ could also affect the correlation as indicated by equation (12). The variation of these parameters may also be responsible for the change of the break frequency on short timescales observed by \cite{gand11}. We noticed, \cite{gall07} suggests that the break to the optically thin portion would take place in the mid-IR (2-40 $\mu$m) for a large sample of BHBs presented in \cite{russ06}, the frequencies of which are smaller than those of H-band ($\sim1.5\ \mu$m), and \cite{gand11} constrained the break frequency of GX 339$-$4 to be $\sim4.6\times10^{13}  {\rm Hz}$, which is also smaller than the frequency of H-band. \cite{CF02} reported a possible jet spectral break in GX 339$-$4 at a frequency around H-band. We guess it is because the spectral break frequency is close to the H-band that leads to the observed break of IR/X correlation in GX 339-4. Small variations in jet parameters will lead to the H-band changing between optically thick and optically thin. Radiation frequencies far below or above the break frequencies will not show a break in the correlation, e.g., the radio/X-ray correlation.

Although the IR/X-ray correlation with the corona or synchrotron origin of X-rays cannot match the steeper branch of the observed one, we cannot rule out the possibility that the X-ray flux originated from the Comptonizing corona or synchrotron radiation of the jet. The magnitude of 3-9 keV flux calculated from equations \eqref{eq6}, \eqref{eq9} and \eqref{eq11} are comparable below the break with the SSC flux a little smaller than that from corona and synchrotron radiation. Above the break, the X-ray flux from SSC is about one order of magnitude higher than that from corona or the synchrotron radiation of the jet. Therefore, the 3-9 keV X-ray flux should be  dominated by the SSC of the jet at low luminosity to match the observed steeper branch with index $\sim 0.68$. This can be achieved by increasing the jet radius $R_0$ or magnetic field $B_0$ at the jet base since the SSC flux is proportional to $R_0^3B_0^6$ from equation \eqref{eq9} and the synchrotron radiation flux is proportional to $R_0^2B_0^4$ from equation \eqref{eq12} for $p=3$ when the accretion rate is low. Therefore, our results suggest that the jet of GX 339$-$4 in the decaying phase (usually with lower luminosity compared to the rising phase) probably has stronger magnetic field or larger jet radius than it is in the rising phase. This is consistent with the work of \cite{barn22}, in which the radio and X-ray flux in the hard states of four outbursts of GX 339$-$4 were investigated based on a model composed of a truncated outer SSD and a inner jet emitting disk. Their results suggest that the radiative or dynamical properties of the jet are different in the rising and decaying phases. The rising phase has weaker magnetic fields and the jet power is dominated by BZ process, while the decaying phase has stronger magnetic fields and the jet power is dominated by BP process. We find that the synchrotron radiation flux is proportional to $\dot{m}^{a_7}$ with $a_7=0.875$ based on equation \eqref{eq15} under the parameters of Figure \ref{fig:3}, which is very close to the results obtained by \cite{barn22} that the radio flux at 9 GHz is proportional to $\dot{m}^{0.9}$ in the decaying phase. 

As shown in Figure 4 of C09, most of the observed data in the steeper and gentler branches correspond to the outburst decaying and rising, respectively. Then the break of the IR/X-ray correlation may also be caused by the change of the jet parameters. If X-ray flux originates from the Comptonizing corona , then for the conical jet with magnetic fields perpendicular to the jet axis ($a_1=a_2=1$), we have $0.68\leqslant a_7/2\leqslant0.71 $ and $0.47\leqslant a_7/2\leqslant0.56 $ for the ballistic and adiabatic jets respectively with $2\leqslant p\leqslant3$ based on equations \eqref{eq11} and \eqref{eq15}, which is in accordance with the indices $\sim0.68$ and $\sim0.48$ of the steeper and gentler branches, respectively.  This implies that the jet changes from adiabatic to ballistic when a BHB undergoes a transition from the outburst rising phase to the decaying phase. The same results can be obtained if the X-ray flux originates from the synchrotron radiation or SSC of the jet. This may be due to the reason that the jet in the rising phase interacts with the surrounding medium and undergoes adiabatic expansion, pushing away the surrounding material, thus forming a ballistic jet during the decaying phase. Further investigations on the jet properties associated with state transitions are expected in the future work.

\textbf{Acknowledgements} This work is supported by National Natural Science Foundation of China under grant Nos. 11803009 and 11603009, and by the Natural Science Foundation of Fujian Province under grant Nos. 2023J01806, 2018J05006, 2018J01416 and 2016J05013.

\bibliography{IRX}

\begin{thebibliography}{41}
\providecommand{\natexlab}[1]{#1}
\providecommand{\url}[1]{{#1}}
\providecommand{\urlprefix}{URL }
\providecommand{\doi}[1]{\url{https://doi.org/#1}}
\providecommand{\eprint}[2][]{\url{#2}}
 \bibcommenthead

\bibitem[{{Band} and {Grindlay}(1985)}]{BG85}
{Band} DL, {Grindlay} JE (1985) {The synchrotron-self-Compton process in
  spherical geometries. I - Theoretical framework}. \apj 298:128--146.
  \doi{10.1086/163593}

\bibitem[{{Barnier} et~al(2022){Barnier}, {Petrucci}, {Ferreira}, {Marcel},
  {Belmont}, {Clavel}, {Corbel}, {Coriat}, {Espinasse}, {Henri}, {Malzac}, and
  {Rodriguez}}]{barn22}
{Barnier} S, {Petrucci} PO, {Ferreira} J, et~al (2022) {Clues on jet behavior
  from simultaneous radio-X-ray fits of GX 339-4}. \aap 657:A11.
  \doi{10.1051/0004-6361/202141182},
  {\href{https://arxiv.org/abs/2109.02895}{{https://arxiv.org/abs/arXiv:2109.02895}}}
  {[astro-ph.HE]}

\bibitem[{{Belloni}(2005)}]{bell05}
{Belloni} T (2005) {Black Hole States: Accretion and Jet Ejection}. In:
  {Burderi} L, {Antonelli} LA, {D'Antona} F, et~al (eds) Interacting Binaries:
  Accretion, Evolution, and Outcomes, American Institute of Physics Conference
  Series, vol 797. AIP, pp 197--204, \doi{10.1063/1.2130233},
  \eprint{astro-ph/0504185}

\bibitem[{{Belloni}(2010)}]{bell10}
{Belloni} TM (2010) {States and Transitions in Black Hole Binaries}. In:
  {Belloni} T (ed) Lecture Notes in Physics, Berlin Springer Verlag, vol 794.
  p~53, \doi{10.1007/978-3-540-76937-8_3}

\bibitem[{{Beloborodov}(1999)}]{belo99}
{Beloborodov} AM (1999) {Plasma Ejection from Magnetic Flares and the X-Ray
  Spectrum of Cygnus X-1}. \apjl 510(2):L123--L126. \doi{10.1086/311810},
  {\href{https://arxiv.org/abs/astro-ph/9809383}{{https://arxiv.org/abs/arXiv:astro-ph/9809383}}}
  {[astro-ph]}

\bibitem[{{Blandford} and {K{\"o}nigl}(1979)}]{BK79}
{Blandford} RD, {K{\"o}nigl} A (1979) {Relativistic jets as compact radio
  sources.} \apj 232:34--48. \doi{10.1086/157262}

\bibitem[{{Blandford} and {Payne}(1982)}]{BP82}
{Blandford} RD, {Payne} DG (1982) {Hydromagnetic flows from accretion disks and
  the production of radio jets.} \mnras 199:883--903.
  \doi{10.1093/mnras/199.4.883}

\bibitem[{{Blandford} and {Znajek}(1977)}]{BZ77}
{Blandford} RD, {Znajek} RL (1977) {Electromagnetic extraction of energy from
  Kerr black holes.} \mnras 179:433--456. \doi{10.1093/mnras/179.3.433}

\bibitem[{{Corbel} and {Fender}(2002)}]{CF02}
{Corbel} S, {Fender} RP (2002) {Near-Infrared Synchrotron Emission from the
  Compact Jet of GX 339-4}. \apjl 573(1):L35--L39. \doi{10.1086/341870},
  {\href{https://arxiv.org/abs/astro-ph/0205402}{{https://arxiv.org/abs/arXiv:astro-ph/0205402}}}
  {[astro-ph]}

\bibitem[{{Corbel} et~al(2003){Corbel}, {Nowak}, {Fender}, {Tzioumis}, and
  {Markoff}}]{corb03}
{Corbel} S, {Nowak} MA, {Fender} RP, et~al (2003) {Radio/X-ray correlation in
  the low/hard state of GX 339-4}. \aap 400:1007--1012.
  \doi{10.1051/0004-6361:20030090},
  {\href{https://arxiv.org/abs/astro-ph/0301436}{{https://arxiv.org/abs/arXiv:astro-ph/0301436}}}
  {[astro-ph]}

\bibitem[{{Corbel} et~al(2013){Corbel}, {Coriat}, {Brocksopp}, {Tzioumis},
  {Fender}, {Tomsick}, {Buxton}, and {Bailyn}}]{corb13}
{Corbel} S, {Coriat} M, {Brocksopp} C, et~al (2013) {The `universal'
  radio/X-ray flux correlation: the case study of the black hole GX 339-4}.
  \mnras 428(3):2500--2515. \doi{10.1093/mnras/sts215},
  {\href{https://arxiv.org/abs/1211.1600}{{https://arxiv.org/abs/arXiv:1211.1600}}}
  {[astro-ph.HE]}

\bibitem[{{Coriat} et~al(2009){Coriat}, {Corbel}, {Buxton}, {Bailyn},
  {Tomsick}, {K{\"o}rding}, and {Kalemci}}]{C09}
{Coriat} M, {Corbel} S, {Buxton} MM, et~al (2009) {The infrared/X-ray
  correlation of GX 339-4: probing hard X-ray emission in accreting black
  holes}. \mnras 400(1):123--133. \doi{10.1111/j.1365-2966.2009.15461.x},
  {\href{https://arxiv.org/abs/0909.3283}{{https://arxiv.org/abs/arXiv:0909.3283}}}
  {[astro-ph.HE]}

\bibitem[{{Dexter} and {Begelman}(2024)}]{DB24}
{Dexter} J, {Begelman} MC (2024) {A relativistic outflow model of the X-ray
  polarization in Cyg X-1}. \mnras 528(1):L157--L160.
  \doi{10.1093/mnrasl/slad182},
  {\href{https://arxiv.org/abs/2308.01963}{{https://arxiv.org/abs/arXiv:2308.01963}}}
  {[astro-ph.HE]}

\bibitem[{{Fender} et~al(2004){Fender}, {Belloni}, and {Gallo}}]{FBG04}
{Fender} RP, {Belloni} TM, {Gallo} E (2004) {Towards a unified model for black
  hole X-ray binary jets}. \mnras 355(4):1105--1118.
  \doi{10.1111/j.1365-2966.2004.08384.x},
  {\href{https://arxiv.org/abs/astro-ph/0409360}{{https://arxiv.org/abs/arXiv:astro-ph/0409360}}}
  {[astro-ph]}

\bibitem[{{Gallo} et~al(2003){Gallo}, {Fender}, and {Pooley}}]{GFP03}
{Gallo} E, {Fender} RP, {Pooley} GG (2003) {A universal radio-X-ray correlation
  in low/hard state black hole binaries}. \mnras 344(1):60--72.
  \doi{10.1046/j.1365-8711.2003.06791.x},
  {\href{https://arxiv.org/abs/astro-ph/0305231}{{https://arxiv.org/abs/arXiv:astro-ph/0305231}}}
  {[astro-ph]}

\bibitem[{{Gallo} et~al(2007){Gallo}, {Migliari}, {Markoff}, {Tomsick},
  {Bailyn}, {Berta}, {Fender}, and {Miller-Jones}}]{gall07}
{Gallo} E, {Migliari} S, {Markoff} S, et~al (2007) {The Spectral Energy
  Distribution of Quiescent Black Hole X-Ray Binaries: New Constraints from
  Spitzer}. \apj 670(1):600--609. \doi{10.1086/521524},
  {\href{https://arxiv.org/abs/0707.0028}{{https://arxiv.org/abs/arXiv:0707.0028}}}
  {[astro-ph]}

\bibitem[{{Gandhi} et~al(2011){Gandhi}, {Blain}, {Russell}, {Casella},
  {Malzac}, {Corbel}, {D'Avanzo}, {Lewis}, {Markoff}, {Cadolle Bel}, {Goldoni},
  {Wachter}, {Khangulyan}, and {Mainzer}}]{gand11}
{Gandhi} P, {Blain} AW, {Russell} DM, et~al (2011) {A Variable Mid-infrared
  Synchrotron Break Associated with the Compact Jet in GX 339-4}. \apjl
  740(1):L13. \doi{10.1088/2041-8205/740/1/L13},
  {\href{https://arxiv.org/abs/1109.4143}{{https://arxiv.org/abs/arXiv:1109.4143}}}
  {[astro-ph.HE]}

\bibitem[{{Georganopoulos} and {Marscher}(1998)}]{GM98}
{Georganopoulos} M, {Marscher} AP (1998) {A Viewing Angle-Kinetic Luminosity
  Unification Scheme for BL Lacertae Objects}. \apj 506(2):621--636.
  \doi{10.1086/306273},
  {\href{https://arxiv.org/abs/astro-ph/9806170}{{https://arxiv.org/abs/arXiv:astro-ph/9806170}}}
  {[astro-ph]}

\bibitem[{{Hjellming} and {Johnston}(1988)}]{HJ88}
{Hjellming} RM, {Johnston} KJ (1988) {Radio Emission from Conical Jets
  Associated with X-Ray Binaries}. \apj 328:600. \doi{10.1086/166318}

\bibitem[{{Homan} et~al(2005){Homan}, {Buxton}, {Markoff}, {Bailyn}, {Nespoli},
  and {Belloni}}]{homa05}
{Homan} J, {Buxton} M, {Markoff} S, et~al (2005) {Multiwavelength Observations
  of the 2002 Outburst of GX 339-4: Two Patterns of X-Ray-Optical/Near-Infrared
  Behavior}. \apj 624(1):295--306. \doi{10.1086/428722},
  {\href{https://arxiv.org/abs/astro-ph/0501349}{{https://arxiv.org/abs/arXiv:astro-ph/0501349}}}
  {[astro-ph]}

\bibitem[{{Hynes} et~al(2003){Hynes}, {Steeghs}, {Casares}, {Charles}, and
  {O'Brien}}]{hyne03}
{Hynes} RI, {Steeghs} D, {Casares} J, et~al (2003) {Dynamical Evidence for a
  Black Hole in GX 339-4}. \apjl 583(2):L95--L98. \doi{10.1086/368108},
  {\href{https://arxiv.org/abs/astro-ph/0301127}{{https://arxiv.org/abs/arXiv:astro-ph/0301127}}}
  {[astro-ph]}

\bibitem[{{Jones}(1968)}]{jone68}
{Jones} FC (1968) {Calculated Spectrum of Inverse-Compton-Scattered Photons}.
  Physical Review 167(5):1159--1169. \doi{10.1103/PhysRev.167.1159}

\bibitem[{{Kaiser}(2006)}]{kais06}
{Kaiser} CR (2006) {The flat synchrotron spectra of partially self-absorbed
  jets revisited}. \mnras 367(3):1083--1094.
  \doi{10.1111/j.1365-2966.2006.10030.x},
  {\href{https://arxiv.org/abs/astro-ph/0601103}{{https://arxiv.org/abs/arXiv:astro-ph/0601103}}}
  {[astro-ph]}

\bibitem[{{Konigl}(1981)}]{koni81}
{Konigl} A (1981) {Relativistic jets as X-ray and gamma-ray sources.} \apj
  243:700--709. \doi{10.1086/158638}

\bibitem[{{K{\"o}rding} et~al(2006){K{\"o}rding}, {Fender}, and
  {Migliari}}]{kord06}
{K{\"o}rding} EG, {Fender} RP, {Migliari} S (2006) {Jet-dominated advective
  systems: radio and X-ray luminosity dependence on the accretion rate}. \mnras
  369(3):1451--1458. \doi{10.1111/j.1365-2966.2006.10383.x},
  {\href{https://arxiv.org/abs/astro-ph/0603731}{{https://arxiv.org/abs/arXiv:astro-ph/0603731}}}
  {[astro-ph]}

\bibitem[{{Mahadevan}(1997)}]{maha97}
{Mahadevan} R (1997) {Scaling Laws for Advection-dominated Flows: Applications
  to Low-Luminosity Galactic Nuclei}. \apj 477(2):585--601.
  \doi{10.1086/303727},
  {\href{https://arxiv.org/abs/astro-ph/9609107}{{https://arxiv.org/abs/arXiv:astro-ph/9609107}}}
  {[astro-ph]}

\bibitem[{{Markoff} et~al(2005){Markoff}, {Nowak}, and {Wilms}}]{mark05}
{Markoff} S, {Nowak} MA, {Wilms} J (2005) {Going with the Flow: Can the Base of
  Jets Subsume the Role of Compact Accretion Disk Coronae?} \apj
  635(2):1203--1216. \doi{10.1086/497628},
  {\href{https://arxiv.org/abs/astro-ph/0509028}{{https://arxiv.org/abs/arXiv:astro-ph/0509028}}}
  {[astro-ph]}

\bibitem[{{Marscher}(1980)}]{mars80}
{Marscher} AP (1980) {Relativistic jets and the continuum emission in QSOs.}
  \apj 235:386--391. \doi{10.1086/157642}

\bibitem[{{Moderski} et~al(1997){Moderski}, {Sikora}, and {Lasota}}]{MSL97}
{Moderski} R, {Sikora} M, {Lasota} JP (1997) {On Black Hole Spins and Dichotomy
  of Quasars}. In: {Ostrowski} M, {Sikora} M, {Madejski} G, et~al (eds)
  Relativistic Jets in AGNs, pp 110--116,
  \doi{10.48550/arXiv.astro-ph/9706263}, \eprint{astro-ph/9706263}

\bibitem[{{Narayan} and {Yi}(1995)}]{NY95}
{Narayan} R, {Yi} I (1995) {Advection-dominated Accretion: Underfed Black Holes
  and Neutron Stars}. \apj 452:710. \doi{10.1086/176343},
  {\href{https://arxiv.org/abs/astro-ph/9411059}{{https://arxiv.org/abs/arXiv:astro-ph/9411059}}}
  {[astro-ph]}

\bibitem[{{Pe'er} and {Casella}(2009)}]{PC09}
{Pe'er} A, {Casella} P (2009) {A Model for Emission from Jets in X-Ray
  Binaries: Consequences of a Single Acceleration Episode}. \apj
  699(2):1919--1937. \doi{10.1088/0004-637X/699/2/1919},
  {\href{https://arxiv.org/abs/0902.2892}{{https://arxiv.org/abs/arXiv:0902.2892}}}
  {[astro-ph.HE]}

\bibitem[{{Remillard} and {McClintock}(2006)}]{RM06}
{Remillard} RA, {McClintock} JE (2006) {X-Ray Properties of Black-Hole
  Binaries}. \araa 44(1):49--92. \doi{10.1146/annurev.astro.44.051905.092532},
  {\href{https://arxiv.org/abs/astro-ph/0606352}{{https://arxiv.org/abs/arXiv:astro-ph/0606352}}}
  {[astro-ph]}

\bibitem[{{Reynolds}(1982)}]{reyn82}
{Reynolds} SP (1982) {Theoretical studies of compact radio sources. I -
  Synchrotron radiation from relativistic flows.} \apj 256:13--37.
  \doi{10.1086/159881}

\bibitem[{{Russell} et~al(2006){Russell}, {Fender}, {Hynes}, {Brocksopp},
  {Homan}, {Jonker}, and {Buxton}}]{russ06}
{Russell} DM, {Fender} RP, {Hynes} RI, et~al (2006) {Global
  optical/infrared-X-ray correlations in X-ray binaries: quantifying disc and
  jet contributions}. \mnras 371(3):1334--1350.
  \doi{10.1111/j.1365-2966.2006.10756.x},
  {\href{https://arxiv.org/abs/astro-ph/0606721}{{https://arxiv.org/abs/arXiv:astro-ph/0606721}}}
  {[astro-ph]}

\bibitem[{{Russell} et~al(2007){Russell}, {Maccarone}, {K{\"o}rding}, and
  {Homan}}]{russ07}
{Russell} DM, {Maccarone} TJ, {K{\"o}rding} EG, et~al (2007) {Parallel tracks
  in infrared versus X-ray emission in black hole X-ray transient outbursts: a
  hysteresis effect?} \mnras 379(4):1401--1408.
  \doi{10.1111/j.1365-2966.2007.11996.x},
  {\href{https://arxiv.org/abs/0705.3594}{{https://arxiv.org/abs/arXiv:0705.3594}}}
  {[astro-ph]}

\bibitem[{{Russell} et~al(2013){Russell}, {Markoff}, {Casella}, {Cantrell},
  {Chatterjee}, {Fender}, {Gallo}, {Gandhi}, {Homan}, {Maitra}, {Miller-Jones},
  {O'Brien}, and {Shahbaz}}]{russ13}
{Russell} DM, {Markoff} S, {Casella} P, et~al (2013) {Jet spectral breaks in
  black hole X-ray binaries}. \mnras 429(1):815--832.
  \doi{10.1093/mnras/sts377},
  {\href{https://arxiv.org/abs/1211.1655}{{https://arxiv.org/abs/arXiv:1211.1655}}}
  {[astro-ph.HE]}

\bibitem[{{Rybicki} and {Lightman}(1979)}]{RL79}
{Rybicki} GB, {Lightman} AP (1979) {Radiative processes in astrophysics}

\bibitem[{{Shakura} and {Sunyaev}(1973)}]{SS73}
{Shakura} NI, {Sunyaev} RA (1973) {Black holes in binary systems. Observational
  appearance.} \aap 24:337--355

\bibitem[{{Sincell} and {Krolik}(1994)}]{SK94}
{Sincell} MW, {Krolik} JH (1994) {Relativistic Induced Compton Scattering in
  Synchrotron Self-absorbed Sources}. \apj 430:550. \doi{10.1086/174430}

\bibitem[{{You} et~al(2021){You}, {Tuo}, {Li}, {Wang}, {Zhang}, {Zhang}, {Ge},
  {Luo}, {Liu}, {Yuan}, {Dai}, {Liu}, {Qiao}, {Jin}, {Liu}, {Czerny}, {Wu},
  {Bu}, {Cai}, {Cao}, {Chang}, {Chen}, {Chen}, {Chen}, {Chen}, {Chen}, {Chen},
  {Cui}, {Cui}, {Deng}, {Dong}, {Du}, {Fu}, {Gao}, {Gao}, {Gao}, {Gu}, {Guan},
  {Guo}, {Han}, {Huang}, {Huo}, {Jia}, {Jiang}, {Jiang}, {Jin}, {Jin}, {Kong},
  {Li}, {Li}, {Li}, {Li}, {Li}, {Li}, {Li}, {Li}, {Li}, {Li}, {Li}, {Liang},
  {Liao}, {Liu}, {Liu}, {Liu}, {Liu}, {Liu}, {Lu}, {Lu}, {Lu}, {Luo}, {Luo},
  {Ma}, {Meng}, {Nang}, {Nie}, {Ou}, {Qu}, {Sai}, {Shang}, {Song}, {Song},
  {Sun}, {Tan}, {Tao}, {Wang}, {Wang}, {Wang}, {Wang}, {Wang}, {Wang}, {Wen},
  {Wu}, {Wu}, {Wu}, {Xiao}, {Xiao}, {Xiong}, {Xu}, {Yang}, {Yang}, {Yang},
  {Yi}, {Yin}, {You}, {Zhang}, {Zhang}, {Zhang}, {Zhang}, {Zhang}, {Zhang},
  {Zhang}, {Zhang}, {Zhang}, {Zhang}, {Zhang}, {Zhang}, {Zhang}, {Zhang},
  {Zhang}, {Zhao}, {Zhao}, {Zheng}, {Zhou}, {Zhou}, {Zhu}, and {Zhu}}]{you21}
{You} B, {Tuo} Y, {Li} C, et~al (2021) {Insight-HXMT observations of jet-like
  corona in a black hole X-ray binary MAXI J1820+070}. Nature Communications
  12:1025. \doi{10.1038/s41467-021-21169-5},
  {\href{https://arxiv.org/abs/2102.07602}{{https://arxiv.org/abs/arXiv:2102.07602}}}
  {[astro-ph.HE]}

\bibitem[{{Zdziarski} et~al(2004){Zdziarski}, {Gierli{\'n}ski},
  {Miko{\l}ajewska}, {Wardzi{\'n}ski}, {Smith}, {Harmon}, and
  {Kitamoto}}]{zdzi04}
{Zdziarski} AA, {Gierli{\'n}ski} M, {Miko{\l}ajewska} J, et~al (2004) {GX
  339-4: the distance, state transitions, hysteresis and spectral
  correlations}. \mnras 351(3):791--807.
  \doi{10.1111/j.1365-2966.2004.07830.x},
  {\href{https://arxiv.org/abs/astro-ph/0402380}{{https://arxiv.org/abs/arXiv:astro-ph/0402380}}}
  {[astro-ph]}

\end{thebibliography}

\end{document}